# Universal Mortality Law, Life Expectancy and Immortality

**Mark Ya. Azbel'**

**School of Physics and Astronomy, Tel-Aviv University,**

**Ramat Aviv, 69978 Tel Aviv, Israel[+];**

**Max-Planck-Institute für Festkorperforschung – CNRS,**

**F38042 Grenoble Cedex 9, France.**

## Abstract

Well protected human and laboratory animal populations with abundant resources are evolutionary unprecedented, and their survival far beyond reproductive age may be a byproduct rather than tool of evolution. Physical approach, which takes advantage of their extensively quantified mortality, establishes that its dominant fraction yields the exact law, and suggests its unusual mechanism. The law is universal for all animals, from yeast to humans, despite their drastically different biology and evolution. It predicts that the universal mortality has short memory of the life history, at any age may be reset to its value at a significantly younger age, and mean life expectancy extended (by biologically unprecedented small changes) from its current maximal value to immortality. Mortality change is rapid and stepwise. Demographic data and recent experiments verify these predictions for humans, rats, flies, nematodes and yeast. In particular, mean life expectancy increased 6-fold (to "human" 430 years), with no apparent loss in health and vitality, in nematodes with a small number of perturbed genes and tissues. Universality allows one to study unusual mortality mechanism and the ways to immortality.

Keywords:

## Introduction: mortality and physics.

Every new field in physics introduced unanticipated concepts and laws. Even thermodynamics of classical particles with reversible mechanics yielded irreversibility. However, biophysicists reduce complex live systems to conventional models. Consider an alternative approach.

Biological diversity evolved in evolutionary selection of the fittest via death of the frail. In the wild competition for sparse resources is fierce, and only relatively few genetically fittest animals survive to their evolutionary "goal"- reproduction. There are no evolutionary benefits from genetically programmed death of tiny number of survivors significantly beyond reproductive age. Well protected human and laboratory animal populations with abundant resources are evolutionary unprecedented and "unanticipated". Their accidental extrinsic mortality is low. Living conditions, phenotypes, genotypes, tissues of laboratory animals may be manipulated and become evolutionary unprecedented. Not by chance, the very existence of the maximal lifespan remains biologically controversial. It was suggested by Strehler and Mildvan, 1960, estimated by Azbel, 1996; Olshansky, Carnes, 1997; Vaupel, 1997 and biologically motivated by Carnes, Olshansky, 1997, 2001, Carnes et al, 2003, yet challenged by Vaupel, 1997; Jazwinski et al, 1998; Carey et al, 1992; Curtsinger et al, 2002; Oeppen, Vaupel, 2000; Arantes-Oliveira et al, 2003; Vaupel et al, 2003. Azbel, 1999, 2002a argued that evolutionary unusual nature of protected populations with abundant (medical and biological included) resources called for an alternative approach to their mortality, based on accurate mathematical analysis of data rather than on any conventional concepts and theories or on demographic approximations. The data are extensive. Accurate knowledge of human mortality is important for economics, taxation, insurance, etc. The famous astronomer Halley, 1693 (the discover of the Halley comet) started quantitative mortality studies. He was followed by the great mathematician Euler, 1760. The actuary Gompertz, 1825 presented the first universal law of mortality for human advanced age. Thereafter the search for such law for all animals went on (Pearl, Miner, 1935; Deevey, 1947; Strehler, Mildvan, 1960; Carnes et al, 1996; Olshansky, Carnes, 1997; Carnes, Olshansky 1997, 2001), but inadequate mathematical tools could not unravel its exact formula and/or conditions. To better estimate and forecast mortality, demographers dropped the universal law and developed over 15 mortality approximations (Coale et al, 1993; Lee, Carter, 1993), each of them for a given population, e.g., for the Southern Italy. Evolutionary theories of mortality were presented (Kirkwood, Austad, 2000; Finch, Kirkwood, 2000; Charlesworth, 1994). Physicists numerically tested them (Stauffer, 2002). Meanwhile, accurate formulation of the problem allows for its exact mathematical solution. The resulting law (previously established empirically in Azbel, 1999, 2002a), which is universal for protected population of all animals from single cell yeast to humans, reduces a dominant fraction of their mortality at any age to several biologically explicit parameters. With several percent accuracy it approximates over 3000 male and female mortality curves in 18 developed countries over their entire history (except the years during, and immediately after, wars, epidemics, etc). Properly scaled, the law is verified for flies (Azbel, 1998, 1999, 2002b). It predicts that 1) At any age mortality is as plastic as "infant" (till 1 year for humans, 1 day for flies, etc) mortality, has short memory of the previous life history, and rapidly (within few percent of the life span) adjusts to current living conditions. 2) At least until certain age (80 years for humans), together with $q(0)$, it may be significantly decreased and even eliminated (see Fig. 1A in Azbel 1999 and Fig. 4 in 2002b); 3) Mortality of a homogeneous cohort may be reset to its value at a much younger age.

Demographic data and recent experiments verify all these conclusions.

1) Following unification of East and West Germany, within several years mortality in the East declined toward its levels in the West, especially among elderly, despite 45 years of their different life histories (Vaupel et al, 2003). Dietary restriction resulted in essentially the same robust increase in longevity in rats (Yu et al, 1985) and decrease in mortality in Drosophilas (Mair et al, 2003), whenever it was switched on, i.e. independent of the previous "dietary" life history.

2) Arguably, zero mortality was also observed, but not appreciated. In 2001 Switzerland (Human Mortality Database, 2003-further HMD) only 1 (out of 60,000) girl died at 5, 9, and 10 years; 5 girls died in each age group from 4 to 7 and from 9 to 13 years; 10 or less from 2 till 17 years; no more than 16 from 2 till 26. Statistics is similar in all 1999-2002 Western developed countries (HMD). Such low values of a stochastic quantity strongly suggest its zero value, at least in lower mortality groups. Similarly, only 2 (out of 7500) dietary restricted flies died at 8 days (Mair et al, 2003). Jazwinski et al, 1998 presented the first model which stated that a sufficient augmentation of aging process resulted in a lack of aging, and supported this conclusion with experimental evidence on yeast mortality. Changes in small number of genes and tissues in nematodes C. Elegans (Arantes-Oliveira et al, 2003) increased their mean life span six-fold to 124 days (compared to 20 days of the wild type) with no apparent loss in health and vitality. None of these 1368 nematodes died till 25 days, only 5% died till 40 days (twice the wild life span), and only 15% died in the first 3 months.

3) Mortality of the female cohort, born in 1900 in neutral Norway, after 17 years of age continuously decreased till 40 years, when it reversed to its value at 12 years. Then it little changed till 50 years, and at 59 years restored its value at 17 years, i.e. 42 years younger-see Fig. 1. (The cohort probability $q(x)$ to die at any age x is calculated according to the HMD procedures and data. According to the universal law, such mortality decrease is not dominated by death of the frail which alters the composition of the cohort-see Discussion). Dietary restriction, switched on day 14 (Mair et al, 2003), in 3 days restored Drosophila mortality at 7 days, i.e. 10 days younger.

Thus, under certain conditions, predicted short memory, reversal of mortality to much younger age and its vanishing even in very old age are universally verified in flies, rats, and humans; vanishing and very low mortality in yeast, nematodes (including biologically amended ones), flies, and humans. Such agreement between predictions and mortality data for animals as remote as humans, flies, nematodes, yeast is hardly coincidental. It motivated the following study. The study specifies conditions when the universal law is valid, and suggests experiments which may comprehensively verify it and its implications. Similar experiments may also establish biological nature of universality. Universality of the law implies that its mechanism is common to different species, genotypes, phenotypes, living conditions, life histories, all other factors, and allows one to study it, thus the ways to regulate it (possibly to immortality), in the simplest case of, e.g., single cell yeast. Then physical experiments may relate the mechanism to processes in and genetics of a cell, and yield their microscopic model.

## Materials and Methods

Demographic life tables present millions of accumulated mortality data in many countries over their history (see, e.g., HMD), which increased infant mortality 50-fold and doubled the life expectancy. For males and females, who died in a given country in a given calendar year, the data list, in particular, "period" probabilities q(x) (for survivors to x) and d(x) (for live newborns) to die between the ages x and (x+1) [note that d(0)=q(0)]; the probability $l(x)$ to survive to x for live newborns; the life expectancy e(x) at the age x. The tables also present the data and procedures which allow one to calculate the values of q(x), d(x), $l(x)$, e(x) for human cohorts, which were born in a given calendar year. All these data, together with fly life tables (Carey, 1993), are used as "materials".

Total mortality depends on a multitude of factors which describe all kinds of relevant details about the population and its environment, from conception to the age of death: genotypes, life history; acquired components, even the month of death (Doblhammer, Vaupel, 2001) and the possibility of its being the late onset genetic decease (Partridge, Gems, 2002). Age specific factors are also important. From 1851 to 1900 English female mortality decreased 2.6 times for 10 years old and by 5% for breast fed infants, prevented from contaminated food and water (Wohl, 1983). Crop failure in 1773 Sweden increased infant mortality by 30%, and 3.6 times mortality at 10-14 years, above its value in 1751 (Statistisk Arsbok, 1993). The 1918 flu pandemic in Europe accounted for 25% of all deaths, with 65% of the flu deaths at ages 15 and 40 years. It increased Swedish female mortality threefold at 28 years, but little changed it for newborns and elderly. Strongly tubercular mortality pattern in Japan prior to 1949 (Johnson, 1995) yielded equal infant mortalities in 1947 Japan and 1877 Sweden. During the last century period mortality rates in a given country at 0, 10 and 40 years decreased correspondingly 50, 100, and 10 times. In contrast to such mortality decrease (primarily due to improving living conditions, medical ones included), the difference between period mortality rates at the same age and in the same calendar year in different countries is rarely more than twofold.

To elucidate the relative contribution of current living conditions, i.e. the relative role of nurture vs. nature in human mortality, Fig. 6 presents the change in the period mortality rates in Switzerland (during 125 peaceful years, from 1876 till 2001) and in England (from 1841 till 1998, which include the Victorian 1850-1900 years of contaminated food and water, and two World Wars). Their mortality rate at 10 years decreased more than 100 fold; infant mortalities, mortalities at 40 and 80 years decreased correspondingly 50, 10 and 5 fold. In contrast to such mortality decrease (primarily due to improving living conditions, medical ones included), the difference between Swiss and English mortalities at the same age in the same calendar year is rarely more than twofold. Short mortality memory (see the previous section), combined with strong mortality dependence on living conditions (e.g., in 20 years the probability to survive from 80 to 100 years in Western Europe increased 20-fold-Vaupel et al, 2003; female survivability to 80 years increased in Japan from 6% in 1950 to 73% in 1999), suggest to study period data and their dependence on q(0). Mathematics provides the method which allows one to establish universality and its law in uncontrollable non-stationary and heterogeneous conditions, to determine the relative contribution of different non-universal mechanisms, and to unravel a new unusual mechanism of universal mortality.

## Results

Suppose (see the first section) that under certain conditions a dominant fraction of period mortality in all heterogeneous populations is accurately related to a specific number of parameters. Such conjecture is sufficiently restrictive to yield this universal relation and its conditions. Chose "additive" mortality variables, which are the averages of their values in different population groups of the same age. If the population consists of the groups with the number $N^G(x)$ of survivors to age $x$ (in years for humans, days for flies, etc) in a group $G$, then the total number of survivors $N(x)$ is the sum of $N^G(x)$ over all $G$. If $c_G$ is the ratio of the population and $l^G(x)$ is the survivability to $x$ in the group $G$, then the probability $l(x)$ to survive to $x$ for live newborns is $l(x) = N(x)/N(0) = \Sigma\, c_G\, l^G(x) = <\Sigma\, l^G(x)>$, i.e. the average of $l^G(x)$ over all groups. The most age specific additive variable is $d(x) = l(x) - l(x+1)$. The most time specific additive variable is $d(0)$ which depends on the time from conception to $x=1$ only. Since the probability to die at the age $x$ is $q(x) = l(x)d(x)$ and $l(0) = 1$, so "infant mortality" $q(0)=d(0)$. In the simplest case (which may easily be generalized) of one variable, universality implies that the relation between $d(x)$ and $q(0)$ (here and on d and q denote the period fractions which yield the universal law) is the same as the relation between their values in any of the groups in the interval. So, if $d(x) = f_x[q(0)]$, then $d^G(x) = f_x[q^G(0)]$. Since additive $d(x) = <d^G(x)>$, $q(0) = <q^G(0)>$, so $<f_x[q^G(0)]> = f_x(<q^G(0)>)$. According to a simple property of stochastic variables, if any average of an analytical function is equal to the function of the average, then the function is linear. However, linearity is inconsistent with experimental d(x) vs q(0) -see, e.g., Fig. 1. Thus, universality restricts heterogeneity of q(0) to certain segments (which must be finite to allow for at least some population heterogeneity). This implies piecewise linear dependence: in the j-th interval (denote its population as an "echelon"), j = 1, 2, …, J,

$$d(x) = d_j(x) = a_j(x)q(0) + b_j(x) \text{ when } q_j < q^G(0) < q_{j+1} \tag{1}$$

Thus, the universal law must have singularities at echelon boundaries. When infant mortality $q(0)$ of an echelon reaches its boundary, it homogenizes. Since, by Eq. (1), d(x) at all ages reduce to infant mortality, they simultaneously reach the interval boundary and, together with q(0), homogenize there. (Two such "ultimate" boundaries are well known: $q(x) = 0$ implies that nobody dies at, and $\lambda(x) = 0$ implies that nobody survives beyond, the age x). At any age $d_j(x) = d_{j+1}(x)$ when $q(0) = q_{j+1}$. This reduces all $d_j(x)$ to (J+1) universal functions of x and (J-1) universal constants.

Equation (1) maps on phase equilibrium. Present it as

$$d(x) = Cd^{(j)}(x) + (1-C)d^{(j+1)}(x), \tag{2}$$

where

$$d^{(j)}(x) = a_j(x)q^{(j)} + b_j(x) \tag{3}$$

Then $d^{(j)}(x)$ may be interpreted as the universal "equation of state" of the j-th "phase" and C as its "concentration" determined by $q(0) = d(0)$. An echelon reduces to two, an arbitrary population to (J+1) phases.

Consider an arbitrarily heterogeneous population. Suppose its fractions and the fractions of its infant mortality $q(0)$ in the j-th echelon are correspondingly $c_j$ and $f_j = c_jq_j(0)/q(0)$. Then $d(x) = \Sigma c_j d_j(x)$ reduces to the universal dependence on these parameters and q(0):

$$d(x) = aq(0) + b; \quad a = \Sigma f_j a_j; \quad b = \Sigma c_j b_j; \tag{4}$$

where $0 < c_j, f_j < 1$; $\Sigma c_j = \Sigma f_j = 1$. (Here and on the argument x is skipped in a and b).

Since $d(x) = [l(x) - l(x+1)]$, so $q(x) = [l(x) - l(x+1)]/l(x)$ equals

$$q(x) = d(x)/[1-d(0)-d(1)-\ldots-d(x-1)] \tag{5}$$

In a general case three groups are sufficient to establish the linear segment (4), more groups allow for the verification of mortality universality with age.

By Eq. (4), in a general case the universal law reduces mortality to population specific parameters $c_j, f_j$ and $q(0)$; to species specific constants $q_j$ and functions $a_j, b_j$ of age x. The number of population specific parameters in Eq. (4) depends on the heterogeneity of the population. If it reduces to a single echelon, then, by Eq. (1), $d(x)$ vs $q(0)$ is universal. Suppose a population is distributed at two, e.g., the 1-st and 2-nd, echelons with the concentrations $c_1$ and $c_2 = 1 - c_1$ correspondingly. Then

$$q(0) = c_1 q_1(0) + (1-c_1) q_2(0); \quad d(x) = c_1 d_1(x) + (1-c_1) d_2(x) \tag{6}$$

By Eqs. (1, 4, 6), $q_1(0) = \alpha_1 q(0)$, $q_2(0) = \alpha_2 q(0)$, where

$$c_1 = (b_2-b)/(b_2-b_1); \quad \alpha_1 = (a-a_2)/[c_1(a_1-a_2)]; \quad \alpha_2 = (a_1-a)/[(1-c_1)(a_1-a_2)]. \tag{7}$$

The crossover to the next non-universal segment occurs when, e.g., $q_1(0)$ reaches the intersection $q_2 = (b_2 - b_1)/(a_1 - a_2)$ of the first and second universal segments in Eq. (1). Then $q_1(0) = q_2$ implies, by Eqs. (1) and (4), that

$$d^I(x) = a_1 q^I(0) + b_1 \tag{8}$$

[a subscript $I$ denotes an intersection in Eq. (4)]. By Eq. (1), this intersection falls on the first universal linear segment or its extension. Thus, in all two echelon populations, Eq. (4) crossovers are situated at universal segments, and this universality is the criterion of any such population.

When q(0)=0, then empirical universal law (Azbel, 1999, 2000a,b) extrapolates d(x), and thus, by Eq. (1), $b_1(x)$ to zero when 4<x<80. Complemented with the universality, this is consistent with possibly zero (till certain old age) mortality q(x)=d(x)l(x), thus zero d(x) and $b_1(x)$, demonstrated in Section 1 for humans, flies, drosophila, nematodes, single cell yeast. Higher accuracy may change these "empirical" zeroes into small universal values ("the first case"). Alternatively, if $b_1(x)$=0 prior to a certain age x=x* or until x=x**, then, according to a well known mathematical theorem, either ("the second case") $b_1(x)$ must have a singularity at, e.g., x=x**, or $b_1(x)$=0 at any age also beyond x**. The latter ("third") case implies ultimate immortality.

## Discussion

**Quantitative verification.** Figures 1 and 2 verify piecewise linear Eq. (4) with the examples of empirical d(80) vs. q(0) for Japanese and French females and their piecewise linear approximations. (Note that life tables present the probabilities per 100,000 live newborns rather than per 1 as in the paper and figures). Until ~ 65 years, d(x) decreases when q(0) increases. Beyond ~ 85 years, d(x) increases together with q(0). In between, d(x) exhibits a well pronounced maximum (naturally, smeared by generic fluctuations – see Figs. 1, 2). Consider the origin of such dependence on age. The number d(x) is proportional to the probability for a newborn to survive to x, and then to die before the age (x+1). When living conditions improve, the former probability increases, while the latter one decreases. In young age the probability to survive to x is close to 1, d(x) is dominated by the mortality rate, and thus monotonically decreases together with q(0). For sufficiently old age, low probability to reach x dominates. It increases with improving living conditions, i.e. with decreasing q(0), thus d(x) increases with decreasing

q(0). At an intermediate age, when improving living conditions sufficiently increase survival probability, d(x) increase is replaced with its decrease. Then d(x) has a maximum at a certain value of q(0). Further study may yield the new lowest mortality echelon, which will dominate future mortality and its law, and will yield better statistics in old age. Such echelon may also yield the d(x) maximum at 95 and even more years of age.

Demographic data (HMD) demonstrate that, except for few irregular years, at any time mortality d(x) dependence on q(0) in most developed countries (e. g., 1948–1999 Austria, 1921–1996 Canada, 1921– 2000 Denmark, 1841–1898 England, 1941–2000 Finland, 1899–1897 France, 1956–1999 West Germany, 1906–1998 Italy, 1950–1999 Japan, 1950–1999 Netherlands, 1896–2000 Norway, 1861–2000 Sweden, 1876–2001 Switzerland) is piecewise linear-see examples in Figs. 2,3. The crossovers are situated at universal segments-see Fig. 4. According to the previous section-see, e.g., Eq. (8), this implies that with such accuracy the populations always reduce to two segments only. Then the intersections determine the universal (i.e. the same for all countries, thus for all humans) law, presented in Fig. 4. Any population reduces to the universal law and the echelon fractions.

Figures 1 and 2 verify Eq. (2) with the examples of empirical d(80) vs. q(0) for Japanese and French females and their piecewise linear approximations. (Note that life tables present the probabilities per 100,000 live newborns rather than per 1 as in the paper and figures). Until ~ 65 years, d(x) decreases when q(0) increases. Beyond ~ 85 years, d(x) increases together with q(0). In between, d(x) exhibits a well pronounced maximum (naturally, smeared by generic fluctuations – see Figs. 1, 2). Consider the origin of such dependence on age. The number d(x) is proportional to the probability for a newborn to survive to x, and then to die before the age (x+1). When living conditions improve, the former probability increases, while the latter one decreases. In young age the probability to survive to x is close to 1, d(x) is dominated by the mortality rate, and thus monotonically decreases together with q(0). For sufficiently old age, low probability to reach x dominates. It increases with improving living conditions, i.e. with decreasing q(0), thus d(x) increases with decreasing q(0). At an intermediate age, when improving living conditions sufficiently increase survival probability, d(x) increase is replaced with its decrease. Then d(x) has a maximum at a certain value of q(0). The crossovers at q(0)~0.006 amplify significant declines of old age mortality in the second half of the 20-th century (Wilmoth, Horiuch, 1999; Wilmoth et al, 2000; Tuljapurkar et al, 2000) with its spectacular medical progress. Further study may yield the new lowest mortality echelon, which will dominate future mortality and its law, and will yield better statistics in old age. Such echelon may also yield the d(x) maximum at 95 and even more years of age.

To analyze the accuracy of the results quantitatively, consider the number D(x) of, e.g., Swiss female deaths at a given age x in each calendar year (Fig. 5). At 10 years it decreases from 126 in 1876 to 1 in 2001. At 80 years the number of deaths increases (together with the life expectancy) from 231 to 951. The number of deaths depends on the size of the population, e.g. in 1999 Japan at 80 years it is ~ 13,061 (see Fig. 5). As a function of age, D(x) decreases to a minimum (at 10 years), then increases to a maximum- D(89)=1,469 in 2001 Switzerland, and D(87)=18,338 in 1999 Japan, and rapidly vanishes around 100 years. According to statistics, the corresponding stochastic error is ~$2/D^{1/2}$. At 10 years of age it increases from ~20% in 1976 to ~200% in 2001 Switzerland and leads to large fluctuations in q(10)-see Fig. 6. At 40 years it is ~20%; at 80 years it is ~6% in Switzerland and ~2% in Japan. If demographic fluctuations in mortality are consistent with this (minimal for a stochastic quantity) generic error for a given age, denote the corresponding mortality as "regular". Otherwise, denote it as "irregular". Human mortality is irregular only during, and few years after, major wars, epidemics, food and water contamination, etc.

Empirical study (Azbel, 1999) demonstrates that species as remote from humans as protected populations of flies also yield the universal law (possibly with a different number of echelons due to different developmental stages-see Azbel, 2004a for more details). Most remarkable and challenging is the same (when properly scaled with the species-specific values of $q_j$) dependence of human and fly functions $a_j$, $b_j$ on age x, which suggests its universality for all animals. Homogenization at the crossovers was also verified (Azbel, 1999).

**Qualitative verification.** In a given echelon the universal law reduces the period canonic mortality at any age to the same calendar infant mortality in the same echelon. This specifies demographic observation that infant mortality is a sensitive barometer of mortality at any age, and is consistent with clinical studies (Osmond, Barker, 2001) of human cohorts. Infant mortality depends only on short period of time (for humans, e.g., less than 2 years, from conception to 1 year). So, universal mortality at any age has correspondingly short memory. This implies its correspondingly rapid adjustment to changing living conditions, and a possibility of its reversal to much younger age.

Thus, a crucial implication of the universal law is its plasticity, which is closely related to its short memory. The latter is very explicit in experiments where dietary restriction in rats and flies is switched on (Yu et al, 1985, Mair et al, 2003). However, when dietary restriction changes to full feeding, their longevity remains higher than in the control group of animals fully fed throughout life. Also, when fly temperature was lowered from 27 to 18 degrees or vice versa, the change in mortality, driven by life at previous temperature, persisted in the switched flies compared to the control ones. Such long memory of the life history may be related to sufficiently rapid deterioration in living conditions (e.g., temperature or feeding, mimetics included) of the population, where death of the frail significantly alters genetic composition of the population. (Note that the universal law is valid when infant mortality little changes within a day for flies, a month for rats, a year for humans, but may significantly change within their life span). This calls for comprehensive tests of short mortality memory in, and thus of rapid (compared to life span) mortality adaptation to, changing living conditions. Similar tests may verify a possibility to reverse and reset mortality of a homogeneous cohort to a much younger age.

The most biologically primitive animals which yield universality may elucidate its mechanism and the nature of "adjustment stairs" in the universal "mortality ladder". If, e.g., slow changes to different temperatures or erythromycin concentrations demonstrate short memory and universality in yeast mortality, then universality may be reduced to processes in a cell.

The universal law is independent of the population, its life history and living conditions. So, it must be related to accurate intrinsic response to their change. Thus, within its accuracy total mortality of a well protected population equals intrinsic mortality (this specifies the Carnes, Olshansky, 1997 suggestion that the universal law is valid for intrinsic mortality only), while extrinsic mortality is negligible.

The universal minimal and maximal mortality values in each echelon demonstrate and quantify the boundaries of the "stairs" in the universal "ladder" of mortality dynamics in progressively improving protected conditions-Fig. 7. Flies yield the same (when properly scaled) universal law (Azbel, 1999, 2003). The universal law which is preserved in evolution of species as biologically remote as humans and flies, is arguably a conservation law in biology and evolution (Azbel, 2003).

Thus, mortality in protected populations of humans and flies is dominated by the universal law. It is valid only in evolutionary unprecedented well protected populations, when "infant" (till 1 year for humans, 1 day for flies) mortality q(0) is small (less than 0.15 for humans-this specifies a protected population) and, e.g., for humans relatively little changes from one calendar year to the next (this defines "regular" conditions, in contrast to "irregular" ones during, and immediately after, wars, epidemics, etc). During the life span of a generation infant mortality change may be very large (~50-fold), i.e. very rapid on such time scale.

**Life expectancy and immortality.** Non-zero minimal universal mortality beyond certain age implies universal maximal mean life span (whose upper limit is the maximal individual life span). For humans the universal law extrapolates it to ~100 years (Azbel, 1996, 2003). This agrees with human maximal lifespan, which remains ~120 years since ancient Rome (where birth and death data were mandatory on the tombstones) to present time. This is consistent with Strehler, Mildvan, 1960; Azbel, 1996; Carnes et al, 2003.

The extrapolation of the Japanese piecewise linear dependence to q(0)=0 within its accuracy is consistent with the universal law d(80)=0 (see Fig.5), i.e. zero mortality at (and thus presumably prior to) 80 years. It is also consistent with the dependence of the period life expectancies e(0) at birth and e(80) at 80 years on the period birth mortality q(0) of Japanese females (Azbel, 2003). If nobody dies until 80, then e(0) = 80 + e(80). In fact, the values of e(0) and e(80), extrapolated to q(0)=0, are correspondingly 93 and 16 years. Thus, e(80) + 80 = 96 years is just 3% higher than e(0)=93.

In contrast, a small number of perturbed genes and tissues increased mean life span in nematodes 6-fold (to 430 years in human terms), with no apparent loss in health and vitality. (One wonders how their cumulative damage, e.g., mutation accumulation, is eliminated). Universality implies this must be true in all animals, single cell yeast included; suggests that universal mortality is indeed a disposable evolutionary byproduct, and may be directed to immortality.

The mapping of the universal law onto phase equilibrium suggests its possible mechanism, whose scaling parameters are related to biology and genetics of (possibly specific) cells. One may test this mechanism in the case of, e.g., single cell yeast; verify the universality of its mortality law; if necessary, to refine the law with more parameters and estimate the contribution of non-universal mortality (for humans this may be done with existing life tables); to develop a microscopic model of universal mortality and transformation of a multitude of external and internal parameters into its scaling parameters; to establish the nature of these parameters.

**Universal law and evolutionary theories.** Consider the implications of the universal law for evolutionary theories of mortality (Kirkwood, Austad, 2002; Finch, Kirkwood, 2000;

Charlesworth, 1994). Mortality is an instrument of natural selection. Survival of the fittest and death of the frail depend (besides genetics) on the previous diseases, traumas, natural disasters, etc, i.e. on the entire life history. This is inconsistent with short mortality memory. In the wild, competition for sparse resources is fierce, and only relatively few genetically fittest animals survive to their evolutionary "goal"- reproduction. Even human life expectancy at birth was around 40-45 years just a century ago (e.g., 38.64 years for males in 1876 Switzerland-HMD). Thus, there are no evolutionary benefits from genetically programmed death or longevity of very few survivors to older age. The Kirkwood disposable soma theory relates their mortality to life-history trade-off (in optimal allocation of metabolic resources between somatic maintenance and reproduction). Thus, it also implies life-history mortality dependence, which is inconsistent with short memory. Theories of irreparable cumulative damage (to DNA, cells, tissues and organs) relate mortality to accumulating mutations with late-acting deleterious effects; telomeres; free radicals etc. In a homogeneous cohort such damage implies persistent mortality increase with age, which is inconsistent with mortality reversal and reset to much younger age. Significant mortality decrease and possibly elimination imply that cumulative damage is either negligible or repairable. According to the Williams antagonistic pleiotropy theory, genes with good early effects may be favored by selection, although these genes had bad effects, including senescence and death, at later ages. In pre-reproductive and reproductive age, such genes are beneficial for longevity. Thus, in perfect conditions with abundant resources they do not yield mortality, in agreement with the universal law and experimental data in the Introduction. Finite mortality prior to post-reproductive age may be consistent with a different kind of antagonistic pleiotropy. Genes, which are beneficial for longevity in the wild, may be detrimental for universal mortality, which dominates in evolutionary unprecedented protected conditions but was negligible in the wild. Some genes, related to certain biological characteristics, may at any age be (accidentally) beneficial for longevity ("longevity genes"-see Jazwinsky, 1996; Puca et al, 2001; Atzmon et al, 2003; Nawrot et al, 2004).

Universal law implies an accurate, reversible, rapid, stepwise, intrinsic response of the total mortality to current environment. The response transforms all multiple environmental parameters into the infant mortality only. It is independent of genotypes, phenotypes, their conditions and life history. It suggests a possibility of mortality elimination (till certain age). Together with its other implications, it suggests a new mechanism of mortality in protected populations. Arguably, mortality dynamics manifests a new kind of law which does not reduce to known science in the same way as statistical mechanics of any system does not reduce to mechanics of its constituent particles (the former is always reversible, the latter is not (hence the well known entropy increase).

Similar to mortality, aging beyond reproductive age yields no evolutionary benefits, and may also be related, e.g., to irreparable cumulative damage. However, biologically amended nematods challenge the inevitability of aging also. Arantes-Oliveira et al, 2003, addressed aging by examining level of activity of surviving animals. Their dynamics may be quantified. This may allow one to study its correlation with, and relation to, the universal mortality.

The universal law allows for quantitative partitioning of the total mortality (which specifies biologically motivated partitioning of mortality by Carnes, Olshansky, 1997). For a given country, sex, and calendar year it determines the population fractions in the echelons. The dependence of these fractions on, e.g., life history quantifies the impact of antagonistic pleiotropy. The difference between the total mortality and the universal law may be partitioned into stochastic fluctuations (which yield the well known Gaussian distribution), "irregular" fluctuations (related to, e.g., 1918 flu pandemic in Europe; World Wars etc), and systematic deviations (related, in particular, to evolutionary mechanisms). Survivors to very old age are more robust, their mortality is more related to genetics, mutation accumulation and other cumulative irreparable damage (Azbel, 1996). This increases the deviation from the universal law- e.g, for Japanese females in Fig. 2 from ~2% at 60 and 80 years to ~10% at 95 years.

**Conclusions.**

A dominant mechanism of mortality changes with age and living conditions. In the wild natural selection is followed beyond certain age by the antagonistic pleiotropy (including its new kind from the previous section), life-history trade-off, and then, in older age, by mutation accumulation and other irreparable damage. This paper considers populations, whose protection from environment (elements of nature, predators, shortage of resources, etc); social and medical (biological interventions, e.g., heart transplants and artificial hearts, included) help for humans; biological changes (e.g., genetic) in animals were evolutionary unanticipated. When conditions persistently improve, mortality descends down the "mortality ladder", with the spurts at the (simultaneous for all ages) edges of its successive "rungs" (see examples in Figs. 2 and 3), which reduce to a new biological concept of population "echelons" and to the universal law (Fig. 4). The latter, when. properly scaled, is universal, its mechanism is common to species as, and even more, remote as humans and flies. Between the spurts mortality has short memory, i.e. the impact of life history is lost. This implies the possibility of mortality reversal to a much younger age, and of its rapid (compared to the life span) stepwise adjustment to changing conditions. The possibility of significant mortality decrease, and its elimination (with relatively minor biological amendments), implies that cumulative damage may be negligible. Universal mortality law, its spurts and all other biologically implications suggest that its mechanism is common to very different animals, thus biologically non-specific, and may be universally regulated, together with mortality and even certain aspects of aging (possibly with a non-specific "longevity pill"). The previous section suggests comprehensive experimental study of the universal law and its implications. If the study demonstrates that the law is valid in yeast also, the universal mechanism of mortality may be reduced to processes in a cell, and suggest the ways to direct mortality dynamics to successive stepwise life extension. A cartoon of the human mortality ladder is presented in Fig. 5.

The universal law reduces dynamics of mortality in any (arbitrarily heterogeneous) population of species at least as remote as humans and flies to few parameters only. In contrast, even thermodynamics of arbitrarily heterogeneous liquids does not yield a universal law. Combined with other implications of the reversible universal law, this suggests that its theory calls for fundamentally new concepts and approach.

**Acknowledgments**

I am very grateful to I. Kolodnaya for assistance and Fig. 7 cartoon. Financial support from A. vonHumboldt award and R. & J. Meyerhoff chair is highly appreciated.

[+] Permanent address     Phone, fax: +972-3-540-7874  e-mail:azbel@post.tau.ac.il## References

Arantes-Oliveira, A., Berman, J.R., Kenyon, C., (2003), Healthy animals with extreme longevity. Science 302, 611-614

Atzmon G., Rabizadeh, P., Gottesman R., Shuldiner A.R., Barzilai N. Inherited pathway for exceptional longevity. 2003. Am. J. Hum. Genetics 73, 1183B

Azbel' M.Ya., 1996. Universal mortality law and species-specific age. Proc. R. Soc. Lond. B. **263**, 1449-1454

Azbel' M. Ya., 1998, Phenomenological theory of mortality and aging. Physica A249, 472-481

Azbel', M.Ya., 1999. Phenomenological theory of mortality evolution. PNAS USA **96**, 3303-3307.

Azbel', M. Ya., 2002a, The law of universal mortality. Phys. Rev. E 66, 0116107.1-9.

Azbel', M.Ya., 2002b. An exact law can test biological theories of aging. Exp. Geront. **37**, 859-869.

Azbel', M.Ya., 2003. Conservation laws of metabolism and mortality. .Physica A **329**, 436-450

Carey J.R., Liedo P., Orozdo D., Vaupel J. W. 1992. Slowing of mortality rates at older ages in large medfly cohorts, Science 258, 447-461

Carey, J. R., 1993. Applied demography for biologists, Oxford university press, N.Y.

Carnes, B.A., Olshansky, S.J., 1997.A biologically motivated partitioning of mortality. Exp. Gerontol. **32**, 615-631.

**Figure Legends**

Fig. 1. The period probabilities for live newborn 1950-1999 Japanese females to die between 60 and 61 (squares), 80 and 81 (triangles), 95 and 96 (diamonds) years of age vs. infant mortality q(0). Their relative mean squared deviations from their piecewise linear approximations (straight lines) are correspondingly 2.4%, 2.3% and 10%.

Fig. 2. The period probability for live newborn 1898-2001 French females to die between 80 and 81 years of age vs. infant mortality q(0). Its relative mean squared deviation from its piecewise linear approximation (straight lines) is 3.7%. Irregular data for 1914-1918 and 1939-1947 years are disregarded.

Fig. 3. Mortality rates vs. age for the 1900 Norwegian female cohort. (The rates are calculated according to the HMD procedures and data).

Fig. 4. The change in mortality rates q(x) of 1841-1998 English (triangles) and 1876-2001 Swiss (diamonds) females with calendar year. From bottom right to top right: x=10 (black), 40 (empty), 0 (black), 80, 95 (empty signs). The years of 1918 flu pandemic and World Wars in Europe, 1851-1900 years of contaminated food and water in Victorian England are denoted by small signs. Note close mortality rates q(0) and q(80) untill the 20-th century.

Fig. 5. Universal law for d(80) and d(60) (upper and lower curves, thick lines) vs. q(0). Note that d(80) = 0 and d(60) =0 when q(0)=0. Diamonds and squares represent (non-universal) intersections for (from left to right) England (two successive intersections), France, Italy and Japan, Finland, Netherlands, Norway, Denmark, France, England correspondingly. Thin lines extend the universal linear segments.

Fig. 6. Number of female deaths in Switzerland in the first year of life (black diamonds), at 10 (black squares), 40 (white triangles), 80 (empty diamonds) and in Japan at 80 (empty squares) years of age for different calendar years. Large signs denote the 1918 flu pandemic year.

Fig. 7. The ladder of rungs in the human "bridge of death" (a cartoon). Better social and medical protection at its successive rungs implies higher "protective walls" against, thus delay in, death and aging, but does not shift the precipice of the maximal species-specific life span at the bridge end. Biologically unprecedented amendments increase the maximal life span and shift the "bridge of death" end.

**Figure 1**

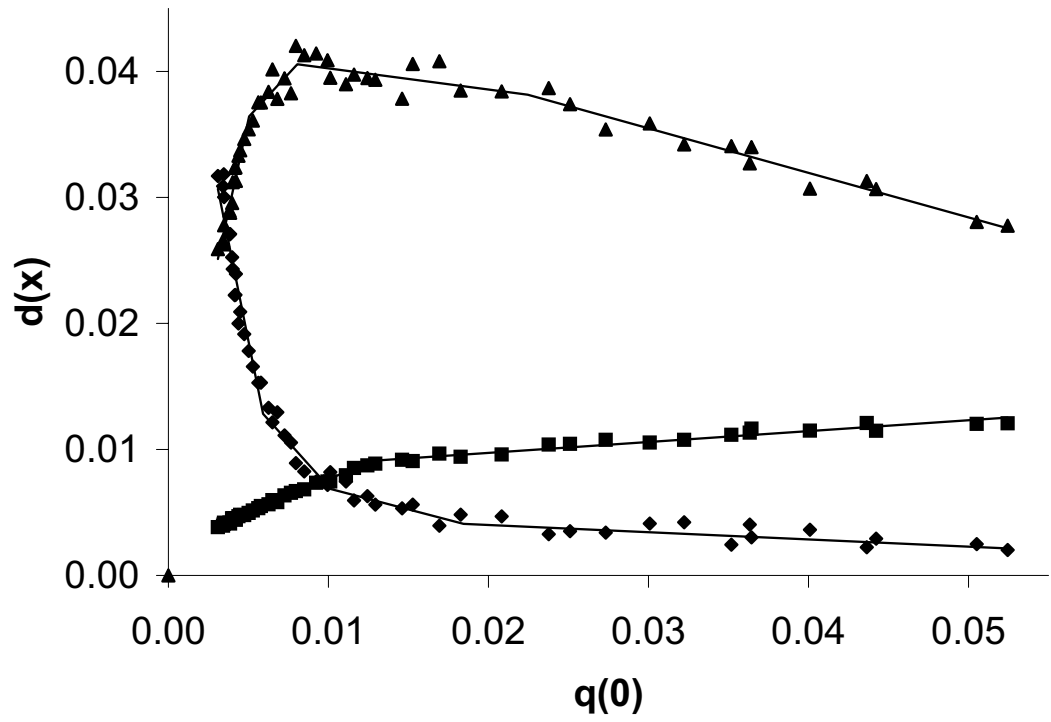

**Figure 2**

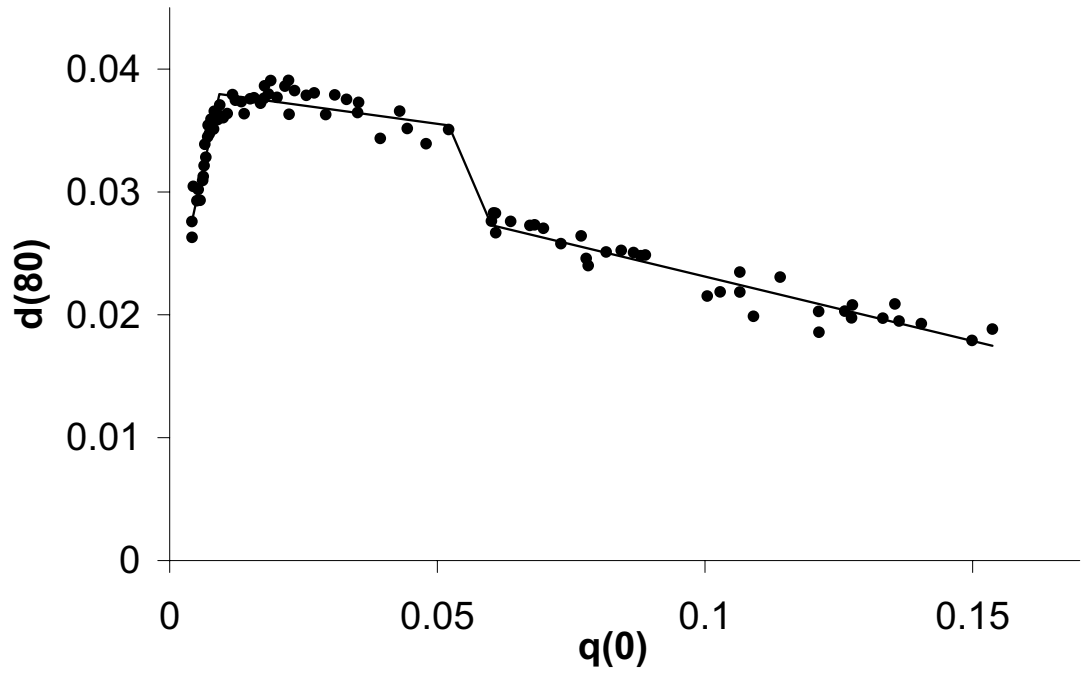

**Figure 3**

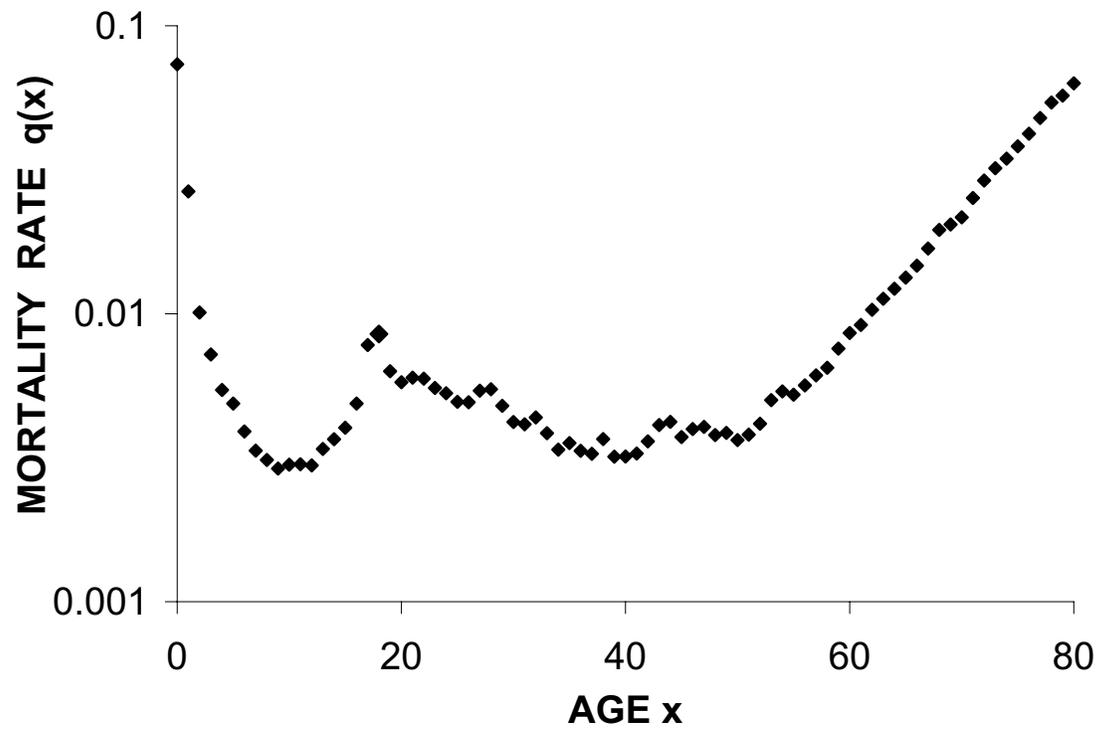

**Figure 4**

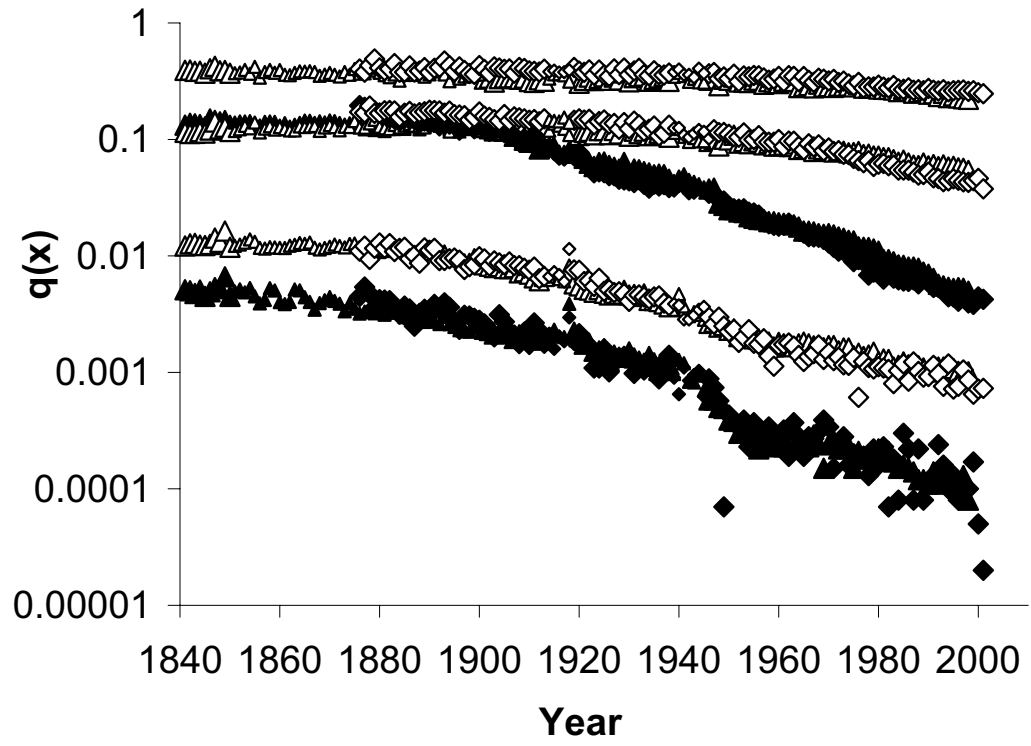

**Figure 5**

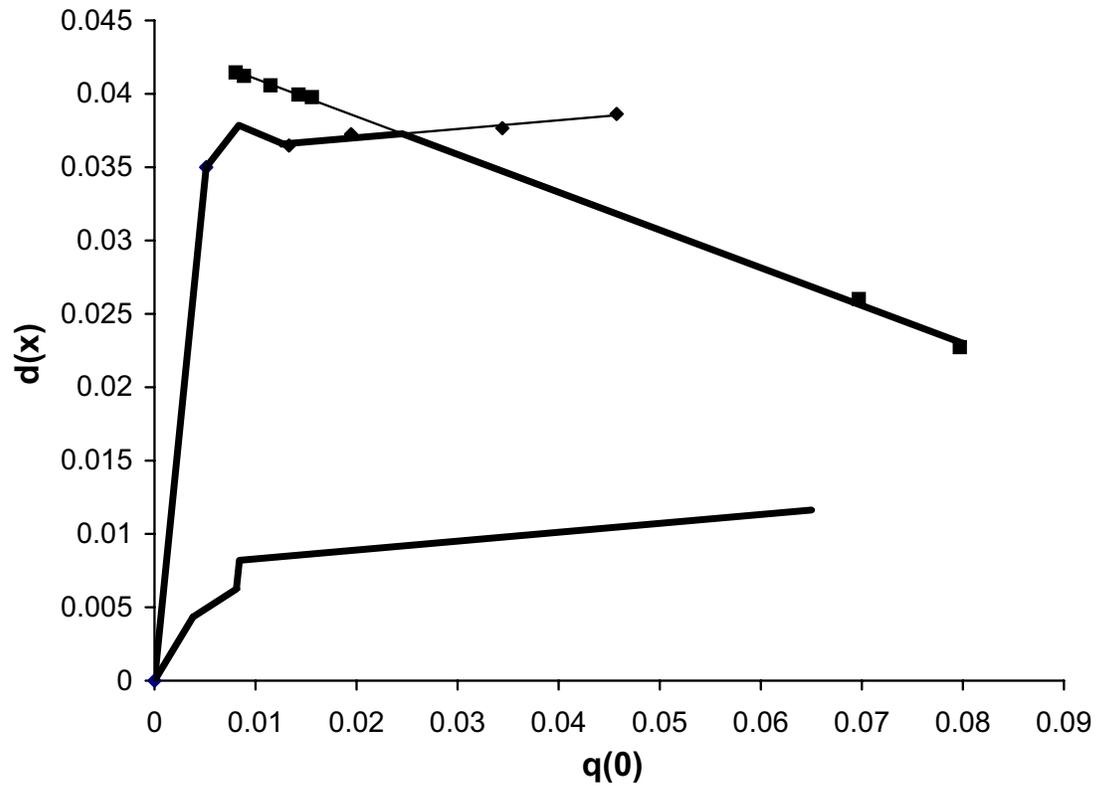

**Figure 6**

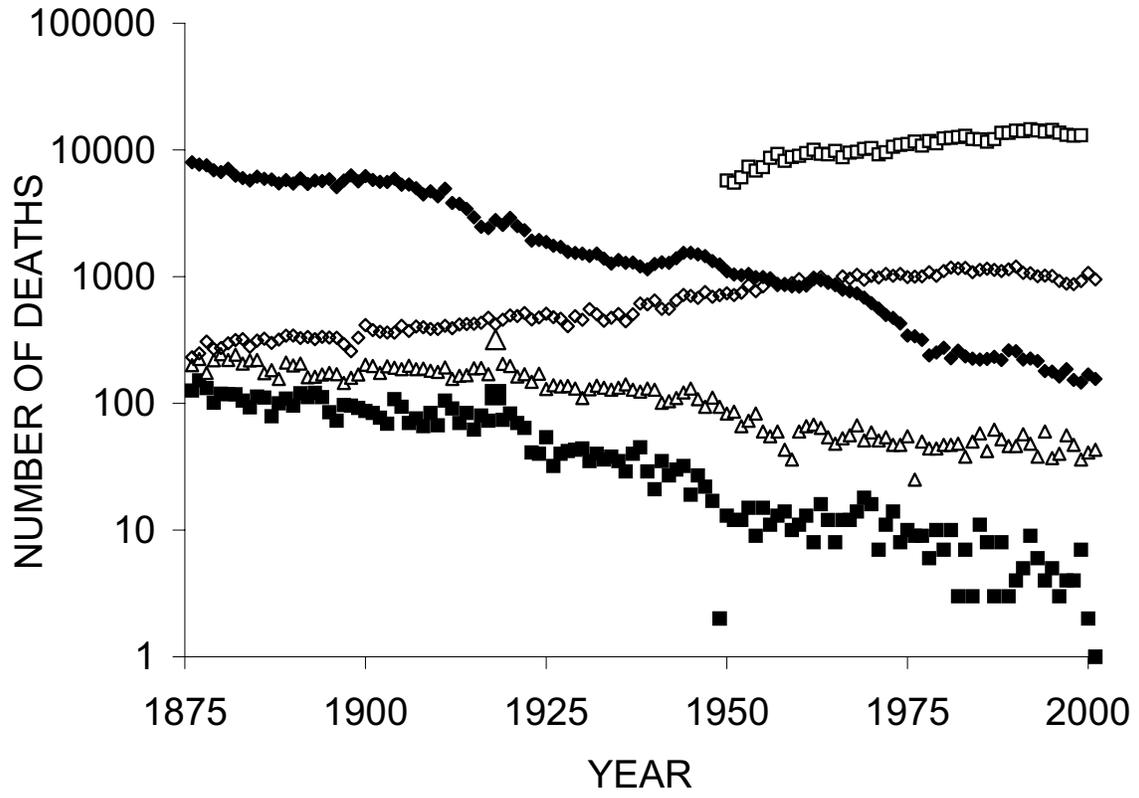

**Figure 7**

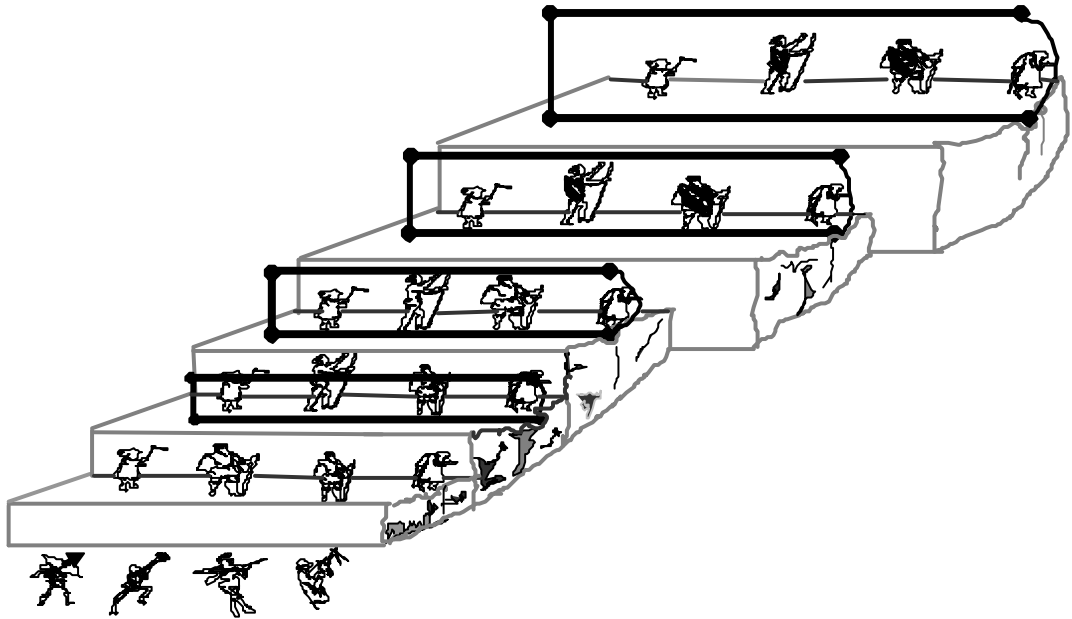